\newcommand{\half}{\frac{1}{2}}
\newcommand{\adag}{a^{\dag}}
\newcommand{\fdag}{f^{\dag}}
\newcommand{\ds}{\displaystyle}
\documentstyle[preprint,aps,tighten]{revtex}
\begin{document}
\draft
\title{q-deformed Fermions}
\author{ P. Narayana Swamy }
\address{Department of Physics, Southern Illinois University,
Edwardsville IL 62026 U.S.A.} \maketitle
\setlength{\oddsidemargin}{0in}

\begin {abstract}

 This is  a study of $q$-Fermions arising from a $q$-deformed
 algebra of harmonic oscillators. Two distinct algebras will be
 investigated.  Employing the first algebra, the Fock states
 are constructed for the   generalized Fermions  obeying Pauli
  exclusion principle. The distribution  function and other
  thermodynamic  properties such as the internal energy and entropy are
  derived. Another  generalization of fermions  from a different
  $q$-deformed algebra is investigated which deals with $q$-fermions not
obeying the exclusion principle. Fock states are constructed for
this system. The basic numbers appropriate for this system are
determined as a direct consequence of the algebra. We also
establish the Jackson Derivative, which is required for the
q-calculus needed to describe these generalized Fermions.

\end{abstract}

\vspace{3.5in}
 Electronic address: pswamy@siue.edu \vspace{.2in}


\section{Introduction}

We shall investigate $q$-deformed Fermions arising as a
consequence of the $q$-deformed algebra of harmonic oscillators.
We shall study two distinct algebras.

First we shall consider generalized Fermions obeying the algebra $
\adag +  q^{-1} \adag a = q^{-N}, \; 0 \leq q \leq 1\, .$ These
generalized Fermions will be shown to obey the exclusion
principle, with the Fock states restricted to $n=0,1$ only. We
shall also see that this algebra is not associated with basic
numbers. We shall investigate in detail, the statistical
thermodynamics of these Fermions which require the use of ordinary
derivatives rather than the Jackson Derivative (JD) of
$q$-calculus. It will be shown that despite the fact that they
obey the exclusion principle, the thermodynamic properties are
quite different from that of ordinary Fermions.

We shall also investigate $q$-deformed Fermions arising from  the
oscillator algebra $a \adag +  q \, \adag a = q^{-N}, \;  0 \leq q
\leq 1\, .$ It will be shown that these generalized Fermions do
not obey the exclusion principle and the Fock states consist of
$n=0,1,2,3, \cdots$ with arbitrary number of quanta. We shall not
investigate the thermodynamics of these Fermions,  and we shall
confine ourselves to a study of the Fock states and some general
properties. We shall establish the JD needed for the $q$-calculus
governing  this system.

Although this investigation, together with a corresponding
$q$-deformed Bosons, accommodates an interpretation as an
interpolating statistics such as in \cite{RAPNS1}, we shall
strictly regard this  as a study of generalized Fermions. The
first algebra investigated here is similar to what is discussed in
an earlier work \cite{ALPNS2} but there are significant
differences as we shall show.

\section{$q$-Fermions obeying Exclusion principle}

Let us begin with the  algebra defined by
\begin{equation}\label{1}
a \adag +  q^{-1} \adag a = q^{-N}, \quad 0 \leq q \leq 1\, ,
\end{equation}
together with  relations
\begin{equation}\label{2}
[N,a]=-a, \quad [N, \adag ]= \adag \, .
\end{equation}
 where $a, \adag$ are the annihilation and creation operators,
 $N$ is the number operator and $q$ is  the deformation parameter, which
 is a
  c-number. This reduces to the standard Fermi algebra in the limit
   $q\rightarrow 1$. To proceed further, let us introduce the
   operator $\adag
a = \hat{N}$, with the concomitant action on Fock states, $\hat{N}
|n \rangle = \beta_n |n \rangle$,  where the eigenvalue depends on
$n$. The relation  $\hat{N} \adag + q^{-1}\adag \hat{N}= \adag
q^{-N}$ follows from the algebra, Eq.(\ref{1}). We may set  $a |n
\rangle = C_n |n-1\rangle; \quad \adag |n\rangle = C'_n
|n+1\rangle \, $, where the constants $C_n, C'_n$ can be
determined.   As a consequence we immediately obtain
\begin{equation}\label{3}
    \beta_{n+1} = q^{-n} - q^{-1} \beta_n \, .
\end{equation}
This recurrence relation can be solved in order to determine
$\beta_n$. We may choose $\beta_0 =0$, thus defining the ground
state as vacuum, i.e., $\adag a |\,0\, \rangle =\beta_0|\,0\,
\rangle =0$.
   We accordingly obtain the solution
 \begin{equation}\label{4}
    \beta_n = 0,\,  1 ,\, 0 , \,q^{-2}, \,0 ,\, q^{-4}, \; \cdots \,
    \quad = \frac{1 - (-1)^n}{2}\, q^{-n+1} \, ,
\end{equation}
which reduces to $\beta_n = 0, \, q^{-n+1}$ respectively when $n$
is an even or odd number. The action of the creation and
annihilation operators on the Fock states yields the results
\begin{equation}\label{5}
    \adag |0 \rangle = \sqrt{\beta_1}|1 \rangle = |1 \rangle; \, \quad
\adag \adag |0 \rangle = \sqrt{\beta_1} \sqrt{\beta_2}|2 \rangle
=0 \, ,
\end{equation}
the sequence of states thus terminates and  consequently the Fock
states are $|0 \rangle, \, |1 \rangle $ only. In other words, the
Fock states in general are built from
\begin{equation}\label{5a}
    |n\rangle = \frac{\ds (\adag)^n}{\ds \sqrt{\beta_n} !}\,
|0\rangle,
\end{equation}
 but restricted to $ n=0,1$ only. The generalized Fermions thus
obey Pauli exclusion principle, just as ordinary Fermions do.

Since $q$-deformed algebras \cite{Bieden} are in general
accompanied by basic numbers \cite{Exton} (bracket numbers) it is
important to stress that there are no basic numbers associated
with this deformed algebra. We note that $\adag a \neq [N]_q,
\quad a \adag \neq [N+1]_q$. Instead, our analysis reveals the
operator relations
\begin{equation}\label{6}
    \adag a = \hat{N}= \frac{1 - (-1)^N}{2}\, q^{-N+1}
    , \quad a \adag = q^{-N} - q^{-1}\hat{N}\, .
    \end{equation}

As this algebra is not related to basic numbers, this formulation
 of $q$-fermions does not require the use of JD and
accordingly we would employ the ordinary derivatives of
thermodynamics rather than that of $q$-calculus. We shall now
proceed to investigate the thermostatistics of these generalized
$q$-Fermions.

  \section{Thermostatics of $q$-fermions}

From the definition of the expectation  value
 \begin{equation}\label{7}
    \hat{n}=
\frac{1}{\cal Z} {\rm Tr} (e^{-\beta H} \hat{N})=
    \frac{1}{\cal Z} {\rm Tr} (e^{-\beta H} \adag a)\, ,
\end{equation}
 and from the form of the Hamiltonian $H = { \sum_{i}} \, N_i \, (E_i - \mu)\,
 ,$ we can determine the distribution function. Using the
 cyclic property of
 the trace and the relations $a f(N) = f(N+1)$, valid for any
polynomial function, we obtain the result
\begin{equation}\label{8}
    {\hat n}_i = \frac{q^{-n_i} }{e^{\beta (E_i - \mu)}+ q^{-1}}\,
    .
\end{equation}
 Employing the result in Eq.(\ref{6}), this may be rewritten as
\begin{equation}\label{9}
    \half  \left ( 1- (-1)^{n}\right )= \frac{q^{-1}}{e^{\beta (E-\mu)} + q^{-1}}\,
    .
\end{equation}
 We may further re-express this result to obtain
the distribution function in the form
\begin{equation}\label{10}
    n_i = \frac{2}{\pi}\arcsin \left
     ( \sqrt{\frac{q^{-1}}{e^{\beta (E_i-\mu)} + q^{-1}} }
    \right )\, ,
\end{equation}
which can then be re-expressed in the form of a power series,
\begin{equation}\label{11}
    n= \frac{1}{\sqrt{g}} + \frac{7 \sqrt{g}}{6} + \frac{149 g^{3/2}}{120}
    + \frac{2161 g^{5/2}}{1680} + \cdots \, ,
\end{equation}
where $ g= q^{-1}/(e^{\eta}+ q^{-1}) $. This form may be used to
determine all the thermodynamic functions for the $q$-fermions.
However,  it is expedient to resort to an approach based on
simplicity which exists due to the exclusion principle.

Recalling that the Fock states reduce to $n=0,1$ only, we observe
that  $\sin^2 n \pi/2 = 0, 1\, $  which can therefore be replaced
by n without losing generality. Consequently the distribution
function reduces to the simple form
\begin{equation}\label{12}
n_i= \frac{q^{-1}}{e^{\beta(E_i-\mu)}+ q^{-1}}\, .
\end{equation}
We may accordingly employ this distribution function following
standard procedure \cite{Huang} to investigate the
thermostatistics of $q$-Fermions, noting that we must employ
ordinary derivatives and not $q$-calculus with Jackson
Derivatives.
  The logarithm of the partition function is
\begin{equation}\label{13}
    \ln {\cal Z}= \sum_i \, \ln (1 + q^{-1}z e^{-\beta E_i})\, ,
\end{equation}
which reproduces the form in Eq.(\ref{12}), namely
\begin{equation}\label{} n_i= z \frac{\partial}{\partial z}=
 \frac{\ds q^{-1}}{\ds e^{\beta(E_i-\mu)}+ q^{-1}}\, .
\end{equation}

Replacing the sum over states by an integration and introducing
the thermal wavelength, $\lambda = h/\sqrt{2 \pi mkT}$ in the
standard manner \cite{Huang}, we determine the  expression for the
thermodynamic potential
\begin{equation}\label{14}
    \Omega = -\frac{1}{\beta}\, \ln {\cal Z} = -\frac{1}{\beta
    \lambda^3}\, \ln (1 + q^{-1}z)
    - \frac{1}{\beta \lambda^3}
    \, f_{5/2}(q,z)\, ,
\end{equation}
where the function $f_n$  defined by
\begin{equation}\label{15}
    f_n(q^{-1}z)= \sum_{r=1}^{\infty}\, (-1)^{r+1}\,
    \frac{(q^{-1}z)^r}{r^n}\,
    ,
\end{equation}
is the  generalized  Riemann Zeta function for $q$-fermions. The
first term in Eq.(\ref{14})  signifies that we have isolated the
zero momentum state in the standard manner.  All of these
thermodynamic functions reduce to  the standard Fermion case in
the limit when $q \rightarrow 1$.

The pressure is determined in the thermodynamic limit:
\begin{equation}\label{16}
    P = \lim_{V \rightarrow \infty, N \rightarrow \infty}
     \left ( - \frac{\Omega}{V} \right )\, =
\frac{1}{\beta \lambda^3}\, f_{5/2}(q^{-1}z)\, ,
\end{equation}
which agrees with the familiar expression in the Fermi limit. The
mean density  in the thermodynamic limit is given by
\begin{equation}\label{17}
\frac{n}{V}= \frac{1}{\lambda^3}\, f_{3/2}(q^{-1}z)\, .
\end{equation}
To determine how these thermodynamic quantities compare with the
corresponding ones for ordinary fermions, we can employ the
familiar graphs
 for the functions $f_{3/2}, f_{5/2}$.  Thus the pressure of the
 $q$-fermions is greater
than that of ordinary fermions at the same temperature and for the
same fugacity. Some of the conclusions agree with earlier work
\cite{ALPNS2}  but there are significant differences. It is
important to stress that the algebra in Eq.(\ref{1}) for
$q$-Fermions  is the same as in ref.\cite{ALPNS2} but here we have
no basic numbers, and no $q$-calculus with JD.

We shall now examine the virial expansion. In the standard
notation, we obtain the result
\begin{equation}\label{18}
    \frac{P v}{k T}= 1 + \frac{1}{2^{5/2}} \left ( \frac{\lambda^3} {v}
     \right )
    + \left ( \frac{1}{8}- \frac{2}{3^{5/2}} \right ) \left
    ( \frac{\lambda^3} {v}\right )^2 +
    \cdots \, .
\end{equation}
It is interesting to note that the virial coefficients are
independent of $q$, hence do not show deformation. Indeed it is
the same as for ordinary fermions and differs from
ref.\cite{ALPNS2}. Furthermore this situation contrasts with the
conclusions of earlier work \cite{RAPNS1} where the formulation
was done in two dimensional space based on an ansatz for the
distribution function. The present investigation is valid in
ordinary 3+1 dimensional space.

The internal energy is given by
\begin{equation}\label{19}
    U= \frac{3 k T V}{2 \lambda^3}f_{5/2}(q^{-1}z)\, .
\end{equation}
 The entropy of the $q$-fermion
systems is determined by the expression
\begin{equation}\label{20}
    \frac{S}{N k}= \frac{5}{2}\, \frac{f_{5/2}(q^{-1}z)}{f_{3/2}(q^{-1}z)}
    - \ln z \, .
\end{equation}
These results possess the expected Fermi limits. We observe that
for $q \neq 1$ the entropy is larger than that of ordinary
fermions. We shall now state some further general results for the
$q$-fermions.

In the limit of large energy,  distribution function reduces to
\begin{equation}\label{21}
    n_i \longrightarrow q^{-1} e^{- \beta E_i}\, ,
\end{equation}
which, other than the normalization factor, reduces to the quantum
Boltzmann statistics. In the limit when $E=\mu$, the distribution
reduces to
\begin{equation}\label{22}
    n_i= \frac{q^{-1}}{1 + q^{-1}} \geq \half \, ,
\end{equation}
which takes the value $\half $ only in the Fermi limit when $q=1$.
In the low temperature limit, when $T \rightarrow 0$, it is clear
from Eq.(\ref{13}) that the distribution function reduces to the
standard unmodified step form for all values of $q$. Hence the
effect of the deformation  may be interpreted  solely as a finite
temperature effect. The modification at higher temperatures is
similar to the standard Fermions except that the parameter $q$
also plays a role.

The dependence on the parameter $q$ is somewhat subtle for many of
the thermodynamic functions and it is worthwhile discussing this.
As an illustration, let us examine the dependence of the
Fermi-energy in some detail. The number density   is given by the
distribution function
\begin{equation}\label{23}
    \frac{N}{V}= \frac{1}{\lambda^3}\, f_{3/2}(q^{-1}z)\, ,
\end{equation}
where, for the sake of simplicity, we have omitted the
multiplicity factor. This can be expressed by the series
\begin{equation}\label{24}
\frac{N}{V}= \frac{4 \pi}{3}\, \left ( \frac{2m k T}{h^2} \right
)^{3/2} \, (\ln (q^{-1}z)^{3/2})\, \left ( 1 + \frac{\pi^2}{8}
(\ln(q^{-1}z)^{-2})+ \cdots \right )\, ,
\end{equation}
and may be employed to determine the chemical potential $\mu$ of
$q$-fermions in terms of the Fermi-energy of standard Fermions,
\begin{equation}\label{25}
    E_F=\frac{3 N}{4 \pi g V}^{2/3}\, \frac{h^2}{2m}\, .
\end{equation}
In the lowest approximation, we obtain
\begin{equation}\label{26}
    \mu = E_F - k T \ln
    {q^{-1}}\, ,
\end{equation}
which shows that the $q$-dependence appears only at finite
temperatures. The expression beyond the zeroth approximation is
given by
\begin{equation}\label{27}
\mu = - k T \ln {q^{-1}} + E_F \left ( 1 - \frac{\pi^2}{12}\,
\left ( \frac{kT}{E_F} \right )^2 + \cdots
 \right )\, .
\end{equation}
Thus the temperature dependence of the chemical potential of
$q$-fermions is different from that of standard Fermions for  $q
\neq 1$.

\section{Parthasarathy-Viswanathan algebra}

We shall now examine the algebra $f \fdag + q \fdag f = q^{-N}\,
$, introduced  by Parthasarathy and Viswanathan \cite{Partha},
together with the relations $[N,f]= -f \; [N,\fdag]=\fdag\, $.
This algebra has also been discussed by Chaichian et al
\cite{Chaichian} , in order to describe fractional statistics.
This algebra reduces to the standard Fermi oscillator algebra in
the limit $q \rightarrow 1$.
 Let the operator ${\tilde
N}= \fdag f \, $ act on the Fock states $|\, n\rangle\, $ so that
${\tilde N}|\, n\rangle = \alpha_n {\tilde N}|\, n\rangle\, $,
where the eigenvalue depends on $n$. The relation
\begin{equation}\label{28}
\quad {\tilde N}\fdag + q \fdag {{\tilde N}}= q^{-N} \fdag \,
\end{equation}
follows directly from the algebra. If we take  $f |\, n\rangle =
C_n |\, n-1\rangle, \quad \fdag |\, n\rangle = C'_n |\, n+1
\rangle\, $, where   $C_n, C'_n$ are constants, we immediately
obtain the relation for any $n$,
\begin{equation}\label{29}
\alpha_{n+1}= q^{-n} - q \, \alpha_n\, .
\end{equation}
 Solving this recurrence relation, we accordingly determine
 $\alpha_n$ to be
 \begin{equation}\label{30}
\alpha_0 =0, \;  \alpha_1=1, \;
 \alpha_2=q^{-1}-q, \; \cdots, \,
 \alpha_n= q^{-n+1}-q^{-n+3}+ \cdots
q^{n-3}-q^{n-1}\, .
\end{equation}
Summing the geometric series, we immediately recognize this as the
basic number,
\begin{equation}\label{31}
 \alpha_n = [n] = \frac{\ds q^{-n}-(-1)^n q^n
}{\ds q + q^{-1}}\, .
\end{equation}
This basic number is characteristic of the algebra and is true for
 $q$-fermions, to be contrasted with a different
 definition applicable for the $q$-Bosons. This is the
 definition introduced by Chaichian
et al and the solution of the recurrence relation above indeed
explains how $f^{\dag}f= [N]$ in terms of the basic number, a
result which is a direct consequence of the algebra. We further
obtain the results
\begin{equation}\label{32}
f |n\rangle = \sqrt{\alpha_n}\; |n-1\rangle, \quad \fdag |n\rangle
= \sqrt{\alpha_{n+1} }\; |n+1 \rangle , \quad  f \fdag = [N+1]\, .
\end{equation}

Let us first examine the Fermi limit when $q \longrightarrow 1$.
We find $\lim_{q\rightarrow 1}\, \alpha_2 = 0$; other $\alpha_n$
may be non-zero in the limit. In the limit, when $q \rightarrow
1$, Fock states are therefore restricted to $|\, 0\rangle , |\,
1\rangle $ only and Pauli exclusion principle is valid only in the
limit. For arbitrary values of $q$, we may construct the Fock
states by
\begin{equation}\label{33}
|n\rangle = \frac{\ds (\adag)^n}{\ds \sqrt{[n]} !}\, |0\rangle\, ,
\end{equation}
where $[n]!= [n] \cdot [n-1]\cdots [1]$.  We observe that this is
different from the rationale provided in ref.\cite{Chaichian} for
invoking the exclusion principle in the limit, hence we need not
assume the relation $f^2= (\fdag)^2=0$.   We also note that
$[n]=\half (1-(-1)^n)$ in the limit $q \rightarrow 1$. As a
consequence, the Fock space breaks up into an infinity of
2-dimensional subspaces when $q=1$, with the Pauli principle valid
in each subspace. On the other hand, $|n\rangle$ exists and is
non-zero for arbitrary $n$ when $q \neq 1$. Consequently these are
generalized fermions with $n=0,1,2, \cdots $.

The Hamiltonian of the generalized Fermions may be taken to be $H
= \half \hbar \omega ( \, [N]- [N+1] \, )$ with $  E = \half \hbar
\omega (\, [n]- [n+1]\, )$, and hence there is no equal spacing
rule for arbitrary $q$. However,  exclusion principle prevails and
$E \rightarrow -\half \hbar \omega, \; + \half \hbar \omega $
 for $n=0,1$ in the limit.

 As discussed in ref.\cite{Chaichian}, the Fock space for
 complex $q$, is quite tricky. When $q=e^{i \pi/m}$,
 we find $[2m]=0$ when $n=2m= 2 \times {\rm
odd}$. And when $n=m=4 k$, we find $[m]=0$. However, this feature
is not present when $q$ is real and therefore the formulation with
real $q$ has distinct advantages.

The basic number here exhibits skew symmetry i.e., $[n]
\longrightarrow \pm [n]$ for $n= $odd, even and this contrasts
with the situation in other algebras. e.g.,
\begin{equation}\label{34}
[n]_B=\frac{q^n -q^{-n}}{q-q^{-1}} \Longrightarrow
[n]_B(q^{-1})=[n]_B(q)\, .
\end{equation}

The thermodynamics of these generalized Fermions would involve
$q$-calculus with JD. However, special care is needed to in order
to identify $q$-calculus with JD in this formalism, due to the
factor $(-1)^N$ in Eq.(\ref{30}). In order to establish the JD for
the $q$-Fermion algebra, we proceed to analyze as follows.

First, we recall the JD in the $q$-boson case:
\begin{equation}\label{35}
{\cal D}f(x)= \frac{1}{x}\, \frac{q^N-q^{-N}}{q-q^{-1}}\, f(x)=
\frac{f(qx) - f(q^{-1}x)}{x(q-q^{-1})}\, ,
\end{equation}
 which reduces to the ordinary derivative in the limit $q\rightarrow 1$.
 In order to study the case of $q$-fermions,
 arising from the algebra  $f \fdag + q \fdag f =
 q^{-N}$, we  may invoke the holomorphy relation
 $f \Longleftrightarrow {\cal D}_x, \; \fdag \Longleftrightarrow
 x$ due to which the algebra implies ${\cal D}_x x + q x {\cal D} =
  q^{-N}$.
It may be useful to recall: $f \Longleftrightarrow
{\partial/\partial x}, \; \fdag \Longleftrightarrow
 x, \quad N=\fdag f \, ,$ leads to  properties \cite{PNS2} such as
\begin{equation}\label{36}
    q^N x = x q^{N+1}; \; q^N x^r = (qx)^r\, ,
\end{equation}
etc.  From the property $\quad [N]x = x[N+1]\quad $, i.e., $ [N]x+
q x {N}=x q^{-N}$,  we infer the solution of ${\cal D}_x x + q x
{\cal D} = q^{-N}$ to be
\begin{equation}\label{37}
    {\cal D}_x= \frac{1}{x}\, \frac{q^{-N}- (-1)^N q^N}{q+q^{-1}}
\end{equation}
as the appropriate JD for $q$-fermions. If we now employ the
properties
\begin{eqnarray}
   q^N f(x) &= &f(qx),\nonumber \\
  q^{-N} f(x) &=& f(q^{-1}x), \nonumber\\
  (-q)^N f(x) &=& f(-qx)\, ,
\end{eqnarray}  \label{38}
  this can be expressed as a differential operator in the standard manner,
\begin{equation}\label{39}
{\cal D}_x \; f(x)= \frac{1}{x}\; \frac{f(q^{-1}x)-
f(-qx)}{q+q^{-1}}\, ,
\end{equation}
valid for $q$-fermions. This may be contrasted with the JD used in
$q$-boson calculus, where
\begin{equation}\label{40}
{\cal D}_x \; f(x)= \frac{1}{x}\; \frac{f(qx)-
f(q^{-1}x)}{q-q^{-1}}\, .
\end{equation}
For the Fermion case, one can investigate many of the properties
satisfied by the JD. In particular the $q$-Fermion JD  reduces to
the ordinary derivative when $q \rightarrow 1$: that is,
$\lim_{q\rightarrow 1} {\cal D}_x f(x) = f'(x)$ which is easily
established using L'Hospital rule.

{\section{Summary and Conclusion}

We have investigated the consequences of the $q$-deformed algebra
$\adag + q^{-1} \adag a = q^{-N}, \; 0 \leq q \leq 1\, $
describing generalized Fermions which obey the exclusion
principle.  In addition to presenting the mathematical
formulation, we have considered detailed physical applications of
 the
generalized Fermions and obtained the various thermodynamic
functions so that we can see what the physical consequences are.
The algebra, together with all the thermodynamic consequences of
the system of $q$-deformed Fermions reduce to those of ordinary
Fermions in the limit $q \rightarrow 1$. We have determined the
thermodynamic functions such as the partition function, pressure,
and the entropy. We have also determined the dependence on $q$ of
the chemical potential as a function of temperature. This is an
example where the deformation is seen to be a finite temperature
effect.

 The Fock states are constructed by $|n\rangle
 = (\adag)^n/\sqrt{\beta_n}\, !|0\rangle \,  $, where
 $\beta_n$ depends on $q$ and $\beta_n=0,1$ for $n=0,1$.
  The thermodynamic properties of these Fermions are dependent on
 the deformation parameter. However, the algebra nevertheless
  has no basic numbers
 associated with it and the system
 is governed by the ordinary calculus of thermodynamics and not the
 $q$-calculus in
 terms of the JD, in contrast to the earlier work cited \cite{ALPNS2}.
The $q$-Boson algebra corresponds to the basic number $[n]=(q^n -
q^{-n})/(q-q^{-1})$ but the $q$-fermions are not associated with
any basic number. This enables us to understand how the
thermodynamic properties of the generalized Fermions could be
different from systems governed by basic numbers and by JD.

We have also investigated the generalized Fermions, not obeying
the exclusion principle, stemming from a
 $q$-deformed oscillator algebra. We have
established the following basic premises. The Fock states of these
generalized Fermions can be built from the action of the creation
operators and require the use of Fermion basic numbers which
follow directly from the algebra. The q-calculus needed to study
the thermostatics of these Fermions must employ a JD which is
characteristic of the nature of the generalized Fermions. We have
determined the form of this JD which reduces to the ordinary
derivative in the Fermi limit, $q \rightarrow 1$. The basic
numbers as well as the JD occurring in the formulation of these
generalized Fermions are quite distinct from corresponding results
for the $q$-deformed Bosons known in the literature.

\acknowledgments

The author would like to thank A. Lavagno and A. Scarfone of
Politecnico di Torino, Torino, Italy, for valuable discussions on
the subject of $q$-deformed algebras.


\begin{thebibliography}{30}
\bibitem{RAPNS1}
R. Acharya and P. Narayana Swamy, \emph{J. Phys. A:Math. Gen.}
\textbf{27}(1994) 7247-7263.
\bibitem{ALPNS2}
 A. Lavagno and P. Narayana Swamy, \emph{Phys.
Rev. }\textbf{E 65} (2002) 036101.
\bibitem{Bieden}
The earliest investigations on deformed algebras: L.Biedenharn,
\emph{J. Phys. A} \textbf{22} (1989) L873; A. Macfarlane,
\emph{ibid} \textbf{22} (1989)4581.
\bibitem{Exton}
 H.Exton, \emph{q-Hypergeometric functions and
applications} (1983), Ellis Horwood Publishers, Chichester.
\bibitem{Huang} K. Huang, \emph{Statistical Mechanics}, pp.177-178, second
edition, (1987) John Wiley \& Sons, New York;  W.Greiner, L.Neise
and H.Stocker, \emph{Thermodynamics and Statistical Mechanics},
Springer-Verlag, New York. L. Reichl, \emph{A modern course in
Statistical Physics} (1998) John Wiley \& Sons, New York;
R.K.Pathria, \emph{Statistical Mechanics}, Pergamon Press (1972)
New York.
\bibitem{Partha}
R.Parthasarathy and K.S. Viswanathan, \emph{J. Phys.} \textbf{A
24} (1991), 613.
\bibitem{Chaichian}M.
Chaichian \emph{et al}, \emph{J. Phys.} \textbf{A 26} (1993),
4017-4034.
\bibitem{PNS2}
P. Narayana Swamy, \emph{Physica }\textbf{A 328} (2003) 145.

\end{thebibliography}
\end{document}